\documentclass[11pt,twoside]{article}
\usepackage{mathrsfs}
\usepackage{amsmath}
\usepackage{amssymb}
\usepackage{graphicx}
\usepackage{subfigure}
\usepackage{appendix}
\usepackage{cite}
\numberwithin{equation}{section}
\newtheorem{definition}{Definition}

\newtheorem{proposition}{Proposition}

\newtheorem{claim}{Claim}

\DeclareMathOperator{\sech}{sech}
\newcommand*{\QEDB}{\hfill\ensuremath{\square}}

\title{ Time-dependent defects in integrable soliton equations}
\author{Baoqiang Xia and Ruguang Zhou
\\
School of Mathematics and Statistics, Jiangsu Normal
University,\\
 Xuzhou, Jiangsu 221116, P. R. China,
 \\
 E-mail address:
xiabaoqiang@126.com; zhouruguang@jsnu.edu.cn
}

\date{}
\begin{document}
\maketitle
\begin{abstract}
We study $(1+1)$-dimensional integrable soliton equations with time-dependent defects located at $x=c(t)$, where $c(t)$ is a function of class $C^1$.
We define the defect condition as a B\"{a}cklund transformation evaluated at $x=c(t)$ in space rather than over the full line.
We show that such a defect condition does not spoil the integrability of the system.
We also study soliton solutions that can meet the defect for the system.
An interesting discovery is that the defect system admits peaked soliton solutions.

\noindent {\bf Keywords:}\quad integrable defect, B\"{a}cklund transformation, soliton equations

\end{abstract}
\newpage

\section{ Introduction}

In recent years, there arose some interest in the study of defects, or impurities, in classical $(1+1)$-dimensional integrable field theories;
see for example \cite{BT1989,GHW2002,CM1995,HMZ2006,BCZ20041,BCZ20042,BCZ2005,CZ2004,CZ2006,CZ2009,Zambon2014,HK2008,Caudrelier2008,CK2015,AD2012,AD20122,Doikou2016,CP2017}
and references therein.
The presence of defects usually spoil the integrability of a system.
An interesting case, on the other hand, is that the defect condition is in form of a B\"{a}cklund transformation (BT)
frozen at the defect location \cite{BCZ20041,BCZ20042,BCZ2005,CZ2004,CZ2006,CZ2009,Zambon2014,HK2008,Caudrelier2008,CK2015,AD2012,AD20122,Doikou2016,CP2017}.
Such a defect condition was found originally by the Lagrangian approach \cite{BCZ20041,BCZ20042,BCZ2005,CZ2004,CZ2006,CZ2009,Zambon2014}
and was proved later to preserve the integrability of a system by showing the
existence of infinite set of conserved quantities and by implementing the classical $r$-matrix method
\cite{HK2008,Caudrelier2008,CK2015,AD2012,AD20122,Doikou2016}.
The solutions, including soliton and finite-gap solutions, were also derived for this type of integrable defect systems \cite{CZ2006,CP2017}.
We note that the current investigations of the integrable defect problems focused mainly on the case of the defect being at a fixed location;
the moving defect problems had received less attention, despite the fact that it was noticed in \cite{BCZ2005} that the defect can move with a constant speed.

The aim of the present paper is to study time-dependent defects in (1+1)-dimensional integrable soliton equations,
including the nonlinear Schr\"{o}dinger (NLS) equation, Korteweg-de Vries (KdV) equation and modified KdV (mKdV) equation
belonging to the Ablowitz-Kaup-Newell-Segur (AKNS) spectral problems \cite{AKNS}.
More precisely, we will consider (1+1)-dimensional integrable soliton equations associated with the AKNS system
in the presence of a defect at time-dependent location $x=c(t)$, where $c(t)$ is a function of class $C^1$.
We define the defect condition as a BT fixed at the defect location $x=c(t)$ in space rather than over the full line.
We show that the resulting defect systems have infinitely many conservation laws.
Furthermore we implement the classical $r$-matrix method to establish the Liouville integrability of the resulting defect systems.
Our results extend the results of \cite{Caudrelier2008,CK2015}
from the situation of the defect being fixed to the situation of the defect moving with time.

In the present paper, we also study soliton solutions for the time-dependent defect systems.
An illustrative example we take is the KdV equation with an integrable defect that moves with a constant speed.
We find that such a defect KdV equation admits peaked soliton (peakon) solutions.
We note that the peakons were first found in the Camassa-Holm (CH) equation \cite{CH1993,FF1981}.
Here it is worth pointing out that peakons for the CH type equations and peakons presented here should be interpreted in two different senses:
the former ones should be interpreted in a suitable weak sense,
while the latter ones should be interpreted in the sense that there is a time-dependent defect; see section 6 of the present paper for details.

The paper is organized as follows.
In section 2, we briefly review the construction of the conservation laws and BTs
for integrable soliton equations belonging to the AKNS spectral problems.
In section 3, we present the time-dependent defect system with the defect condition corresponding to a BT
and show such a defect system admits a Lagrangian description.
In section 4,  we study the integrability of the time-dependent defect system.
In section 5, we generalize the results of section 4 to the case that there are multiple time-dependent defects in an integrable system.
In section 6, we study soliton solutions for the time-dependent defect KdV equation.
Some concluding remarks are drawn in section 7.

\section{AKNS system, conservation laws and BT}

For self-containedness, we start by a brief review of the construction of the conservation laws and the BTs
for integrable soliton equations associated with the AKNS spectral problems.

We consider AKNS spectral problems \cite{AKNS}:
\begin{subequations}
\begin{eqnarray}
\phi_x(x,t,\lambda)=U(x,t,\lambda)\phi(x,t,\lambda),
~~ U=\left( \begin{array}{cc} -i\lambda & u(x,t) \\
 v(x,t) &  i\lambda \\ \end{array} \right),
 \label{lpx}
 \\
\phi_t(x,t,\lambda)=V(x,t,\lambda)\phi(x,t,\lambda),
~~ V=\left( \begin{array}{cc} V_{11} & V_{12} \\
 V_{21} & -V_{11}  \\ \end{array} \right),
 \label{lpt}
\end{eqnarray}
\label{LPxt}
\end{subequations}
where $\lambda$ is a spectral parameter, $\phi=(\phi_{1},~\phi_{2})^T$,
and $V_{jk}$, $j,k=1,2$, are some functions depend on $u(x,t)$, $v(x,t)$ and on the spectral parameter $\lambda$.
The compatibility condition of (\ref{LPxt}), namely
\begin{eqnarray}
U_t-V_x+\left[U,V\right]=0,
\end{eqnarray}
may generate quite a few important integrable nonlinear evolution equations in the soliton theory.
For example, if we consider the reduction $v=\varepsilon u^*$, $\varepsilon=\pm 1$, and take
\begin{eqnarray}
V=\left( \begin{array}{cc} -2i\lambda^2-i\varepsilon|u|^2 & 2\lambda u+iu_x \\
 \varepsilon(2\lambda u^*-iu^*_x) &  2i\lambda^2+i\varepsilon|u|^2  \\ \end{array} \right),
 \label{nlstl}
\end{eqnarray}
we then obtain the celebrated NLS equation
\begin{eqnarray}
iu_t+u_{xx}-2\varepsilon u|u|^2=0, \quad \varepsilon=\pm 1.
\label{NLS}
\end{eqnarray}
If we consider the reduction $v=-1$ and take
\begin{eqnarray}
V=\left( \begin{array}{cc} -4i\lambda^3+2i\lambda u-u_x & 4\lambda^2 u+2i\lambda u_x-2u^2-u_{xx} \\
 -4\lambda^2+8u &  4i\lambda^3-2i\lambda u+u_x  \\ \end{array} \right),
 \label{kdvtl}
\end{eqnarray}
we then obtain the famous KdV equation
\begin{eqnarray}
u_t+u_{xxx}+6uu_{x}=0.
\label{kdv}
\end{eqnarray}
If we consider the reduction $v=-u$ and take
\begin{eqnarray}
V=\left( \begin{array}{cc} -4i\lambda^3+2i\lambda u^2 & 4\lambda^2 u+2i\lambda u_x-2u^3-u_{xx} \\
-4\lambda^2 u+2i\lambda u_x+2u^3+u_{xx} &  4i\lambda^3-2i\lambda u^2  \\ \end{array} \right),
 \label{mkdvtl}
\end{eqnarray}
we then obtain the mKdV equation
\begin{eqnarray}
u_t+u_{xxx}+6u^2u_{x}=0.
\label{mkdv}
\end{eqnarray}

We will assume, in this paper, the fields $u(x,t)$ for the above equations in the bulk are sufficiently smooth and decay as $|x|\rightarrow \infty$
or as $|t|\rightarrow \infty$.

Let $\Gamma=\frac{\phi_2}{\phi_1}$,
then it follows from (\ref{LPxt}) that $\Gamma$ satisfies the following $x$-part and $t$-part Riccati equations
\begin{subequations}
\begin{eqnarray}
\Gamma_x=2i\lambda \Gamma+v-u\Gamma^2,
\label{ricx}
\\
\Gamma_t=V_{21}-2V_{11} \Gamma-V_{12}\Gamma^2.
\label{rict}
\end{eqnarray}
\label{ric}
\end{subequations}
Moreover, we find from (\ref{LPxt}) that
\begin{eqnarray}
\begin{array}{l}
(\ln \phi_1)_x=-i\lambda+u\Gamma,
\\
(\ln \phi_1)_t=V_{11}+V_{12}\Gamma,
\end{array}
\label{lnp}
\end{eqnarray}
which in turn generates the following conservation law
\begin{eqnarray}
\left(u\Gamma\right)_t=\left(V_{11}+V_{12}\Gamma\right)_x.
\label{CL}
\end{eqnarray}
The functions $u\Gamma$ and $V_{11}+V_{12}\Gamma$ in (\ref{CL}) provide the generating functions for the conservation densities and for the associated fluxes, respectively.
We can derive explicit forms of conservation densities by expanding $\Gamma$ in terms of negative powers of $\lambda$.
Indeed, by substituting the expansion
\begin{equation}
\Gamma=\sum_{n=1}^{\infty}\Gamma_n(2i\lambda)^{-n}
\label{oe1}
\end{equation}
into (\ref{ricx})
and by equating the coefficients of powers of $\lambda$, we arrive at
\begin{eqnarray}
\Gamma_{1}=-v, \qquad \Gamma_{2}=-v_x,
\label{w1}
\end{eqnarray}
and the recursion relation:
\begin{eqnarray}
\Gamma_{n+1}=\left(\Gamma_{n}\right)_x+u\sum_{j=1}^{n-1}\Gamma_j\Gamma_{n-j},\quad n\geq 2.
\label{wj}
\end{eqnarray}
Substituting (\ref{oe1}), (\ref{w1}) and (\ref{wj}) into (\ref{CL}) we finally obtain an infinite set of conservation laws.

We now turn to the construction of BTs for the AKNS system.
We consider another copy of the auxiliary problem for $\tilde{\phi}$ with Lax pair
$\tilde{U}$, $\tilde{V}$ defined as in (\ref{LPxt}) with the new potentials $\tilde{u}$, $\tilde{v}$, replacing $u$, $v$.
We assume that the two systems are related by the gauge transformation
\begin{eqnarray}
\tilde{\phi}(x,t,\lambda)=B(x,t,\lambda)\phi(x,t,\lambda).
\label{BT}
\end{eqnarray}
A necessary and sufficient condition for (\ref{BT}) is that the matrix $B(x,t,\lambda)$ satisfies
\begin{subequations}
\begin{eqnarray}
B_x(x,t,\lambda)=\tilde{U}(x,t,\lambda)B(x,t,\lambda)-B(x,t,\lambda)U(x,t,\lambda),
\label{BT1a}
\\
B_t(x,t,\lambda)=\tilde{V}(x,t,\lambda)B(x,t,\lambda)-B(x,t,\lambda)V(x,t,\lambda).
\label{BT1b}
\end{eqnarray}
\label{BT1}
\end{subequations}
Transformation (\ref{BT}) is actually a Darboux transformation (DT) \cite{MS1991}, since it preserves the forms of the Lax pair.
Equation (\ref{BT1}) induces a relation, called a BT \cite{RS2002}, between the potentials $u$, $v$ and $\tilde{u}$, $\tilde{v}$:
\begin{eqnarray}
\mathcal{B}(u,v,\tilde{u},\tilde{v})=0.
\label{BTpotential}
\end{eqnarray}
For example, for the NLS equation (\ref{NLS}), we may take
\begin{eqnarray}
B=I+\frac{1}{2\lambda}\left( \begin{array}{cc} \alpha+i\Omega & -i(\tilde{u}-u)\\
 i\varepsilon (\tilde{u}-u)^* &  \alpha-i\Omega  \\ \end{array} \right),
~~\Omega=\pm \sqrt{\beta^2 +\varepsilon |\tilde{u}-u|^2},
 \label{nlsdm}
\end{eqnarray}
the corresponding BT becomes
\begin{eqnarray}
\begin{split}
\tilde{u}_x-u_x=i\alpha\left(\tilde{u}-u\right)+\Omega\left(\tilde{u}+u\right),
\\
\tilde{u}_t-u_t=-\alpha\left(\tilde{u}_x-u_x\right)+i\Omega \left(\tilde{u}_x+u_x\right)-i\varepsilon(\tilde{u}-u)\left(|\tilde{u}|^2+|u|^2\right),
\end{split}
\label{deNLS}
\end{eqnarray}
where $\alpha$ and $\beta$ are two arbitrary real constants.
For the KdV equation (\ref{kdv}), we may take
\begin{eqnarray}
B=I+i\lambda^{-1}\left( \begin{array}{cc} \frac{1}{2}\sqrt{\beta^2-2(\tilde{u}+u)} & \frac{1}{2}(\tilde{u}+u) \\
 1 &  -\frac{1}{2}\sqrt{\beta^2-2(\tilde{u}+u)} \\ \end{array} \right),
 \label{kdvdm}
\end{eqnarray}
the corresponding BT becomes
\begin{eqnarray}
\begin{split}
\left(\tilde{u}_x+u_x\right)=\left(\tilde{u}-u\right)\sqrt{\beta^2-2(\tilde{u}+u)},
\\
\left(\tilde{u}_t+u_t\right)=-\left(3(\tilde{u}^2-u^2)+(\tilde{u}-u)_{xx}\right)\sqrt{\beta^2-2(\tilde{u}+u)}.
\end{split}
\label{dekdv}
\end{eqnarray}
For the mKdV equation (\ref{mkdv}), we may take
\begin{eqnarray}
B=I+\frac{i}{2}\lambda^{-1}\left( \begin{array}{cc} \sqrt{\beta^2-(\tilde{u}-u)^2} & -(\tilde{u}-u) \\
 -(\tilde{u}-u) &  -\sqrt{\beta^2-(\tilde{u}-u)^2} \\ \end{array} \right),
 \label{mkdvdm}
\end{eqnarray}
the corresponding BT becomes
\begin{eqnarray}
\begin{split}
\left(\tilde{u}_x-u_x\right)=\left(\tilde{u}+u\right)\sqrt{\beta^2-(\tilde{u}-u)^2},
\\
\left(\tilde{u}_t-u_t\right)=-\left(2(\tilde{u}^3+u^3)+(\tilde{u}+u)_{xx}\right)\sqrt{\beta^2-(\tilde{u}-u)^2}.
\end{split}
\label{demkdv}
\end{eqnarray}

\section{Integrable systems with time-dependent defects}

Let $c(t)$ be a function of class $C^1$.
We study integrable equations with a time-dependent defect placing at $x=c(t)$ in space.
We define the defect condition as a BT evaluated at $x=c(t)$.
We show such a defect system admits a Lagrangian description.

\subsection{Time-dependent defect conditions arising from BTs}

We suppose that the auxiliary problem (\ref{LPxt}) exists for $x>c(t)$, while the one for $\tilde{U}$ and $\tilde{V}$ exists for $x<c(t)$.
At the time-dependent position $x=c(t)$, we assume that the two systems are connected via the condition (\ref{BT}) evaluated at $x=c(t)$.

\begin{definition}
A $(1+1)$-dimensional integrable equation with a defect at time-dependent location $x=c(t)$ in space is described by the following internal boundary problem:
\begin{itemize}
\item $u(x,t)$ and $\tilde{u}(x,t)$ satisfy the equation in the bulk for $x>c(t)$ and for $x<c(t)$, respectively;
\item at $x=c(t)$,  $u(c(t),t)$ and $\tilde{u}(c(t),t)$ are connected by a condition corresponding to the BT for $u(x,t)$ and $\tilde{u}(x,t)$. 
\end{itemize}
\end{definition}
For example, the NLS equation with the above defined time-dependent defect reads
\begin{subequations}
\begin{eqnarray}
iu_t+u_{xx}-2\varepsilon u|u|^2=0, \quad \varepsilon=\pm 1, \quad x>c(t),
\label{defectNLSa}
\\
i\tilde{u}_t+\tilde{u}_{xx}-2\varepsilon \tilde{u}|\tilde{u}|^2=0, \quad \varepsilon=\pm 1, \quad x<c(t),
\label{defectNLSb}
\\
\left.\left(\tilde{u}_x-u_x\right)\right|_{x=c(t)}=\left.\left(i\alpha\left(\tilde{u}-u\right)+\Omega\left(\tilde{u}+u\right)\right)\right|_{x=c(t)},
\label{defectNLSc}
\\
\left.\left(\tilde{u}_t-u_t\right)\right|_{x=c(t)}=\left.\left(-\alpha\left(\tilde{u}_x-u_x\right)+i\Omega \left(\tilde{u}_x+u_x\right)-i\varepsilon(\tilde{u}-u)\left(|\tilde{u}|^2+|u|^2\right)\right)\right|_{x=c(t)},
\label{defectNLSd}
\end{eqnarray}
\label{defectNLS}
\end{subequations}
where
$\Omega=\pm \sqrt{\beta^2 +\varepsilon |\tilde{u}-u|^2}$.
The KdV equation with the time-dependent defect reads
\begin{subequations}
\begin{eqnarray}
u_t+u_{xxx}+6uu_{x}=0, \quad x>c(t),
\label{defectkdva}
\\
\tilde{u}_t+\tilde{u}_{xxx}+6\tilde{u}\tilde{u}_{x}=0, \quad x<c(t),
\label{defectkdvb}
\\
\left.\left(\tilde{u}_x+u_x\right)\right|_{x=c(t)}=\left.\left(\tilde{u}-u\right)\sqrt{\beta^2-2(\tilde{u}+u)}\right|_{x=c(t)},
\label{defectkdvc}
\\
\left.\left(\tilde{u}_t+u_t\right)\right|_{x=c(t)}=\left.-\left(3(\tilde{u}^2-u^2)+(\tilde{u}-u)_{xx}\right)\sqrt{\beta^2-2(\tilde{u}+u)}\right|_{x=c(t)}.
\label{defectkdvd}
\end{eqnarray}
\label{defectkdv}
\end{subequations}
The mKdV equation with the time-dependent defect reads
\begin{subequations}
\begin{eqnarray}
u_t+u_{xxx}+6u^2u_{x}=0, \quad x>c(t),
\label{defectmkdva}
\\
\tilde{u}_t+\tilde{u}_{xxx}+6\tilde{u}^2\tilde{u}_{x}=0, \quad x<c(t),
\label{defectmkdvb}
\\
\left.\left(\tilde{u}_x-u_x\right)\right|_{x=c(t)}=\left.\left(\tilde{u}+u\right)\sqrt{\beta^2-(\tilde{u}-u)^2}\right|_{x=c(t)},
\label{defectmkdvc}
\\
\left.\left(\tilde{u}_t-u_t\right)\right|_{x=c(t)}=\left.-\left(2(\tilde{u}^3+u^3)+(\tilde{u}+u)_{xx}\right)\sqrt{\beta^2-(\tilde{u}-u)^2}\right|_{x=c(t)}.
\label{defectmkdvd}
\end{eqnarray}
\label{defectmkdv}
\end{subequations}

\subsection{Lagrangian descriptions for the defect systems}

We now show that the time-dependent defect system also admits a Lagrangian description.
We will fix our ideas on the above mentioned three examples:
the defect NLS equation (\ref{defectNLS}), the defect KdV equation (\ref{defectkdv}) and the defect mKdV equation (\ref{defectmkdv}).

\subsubsection{Lagrangian formulation for the defect NLS equation}

The NLS equation (\ref{NLS}) in the bulk is described by the Lagrangian
\begin{eqnarray}
L=\int_{-\infty}^{\infty}dx \left(\frac{i}{2}(u^*u_t-uu^*_t)-|u_x|^2-\varepsilon |u|^4\right).
\label{LNLSB}
\end{eqnarray}
To describe the defect NLS equation (\ref{defectNLS}), we modify the Lagrangian as follows
\begin{eqnarray}
L=\int_{-\infty}^{c(t)}dx \mathcal{L}(\tilde{u})+D+\int_{c(t)}^{\infty}dx \mathcal{L}(u),
\label{LNLS}
\end{eqnarray}
where
\begin{eqnarray}
\mathcal{L}(u)=\frac{i}{2}(u^*u_t-uu^*_t)-|u_x|^2-\varepsilon |u|^4
\label{LNLSr}
\end{eqnarray}
is the Lagrangian density of the bulk system for $x>c(t)$,
\begin{eqnarray}
\mathcal{L}(\tilde{u})=\frac{i}{2}(\tilde{u}^*\tilde{u}_t-\tilde{u}\tilde{u}^*_t)-|\tilde{u}_x|^2-\varepsilon |\tilde{u}|^4
\label{LNLSl}
\end{eqnarray}
is the Lagrangian density of the bulk system for $x<c(t)$,
\begin{eqnarray}
\begin{split}
D=&-\frac{i}{2}\varepsilon\omega\left(\frac{\dot{\tilde{\mathfrak{u}}}-\dot{\mathfrak{u}}}{\tilde{\mathfrak{u}}-\mathfrak{u}}-
\frac{\dot{\tilde{\mathfrak{u}}}^*-\dot{\mathfrak{u}}^*}{\tilde{\mathfrak{u}}^*-\mathfrak{u}^*}\right)-\frac{1}{3}\varepsilon\omega^3
+\omega\left(|\tilde{\mathfrak{u}}|^2+|\mathfrak{u}|^2+\varepsilon \alpha^2-\varepsilon\alpha c'(t)\right)
\\&+(\frac{i}{2}c'(t)-i\alpha)\left(\tilde{\mathfrak{u}}^*\mathfrak{u}-\tilde{\mathfrak{u}}\mathfrak{u}^*\right)
\end{split}
\label{D}
\end{eqnarray}
is the defect contribution at $x=c(t)$.
In (\ref{D}), we used the following abbreviated expression
\begin{subequations}
\begin{eqnarray}
\mathfrak{u}=u(c(t),t), ~~\tilde{\mathfrak{u}}=\tilde{u}(c(t),t), ~~\omega=\pm \sqrt{\beta^2 +\varepsilon |\tilde{\mathfrak{u}}-\mathfrak{u}|^2},
\label{DEa}
\\
\mathfrak{u}_1=u_x(x,t)\mid_{x=c(t)}, ~~\tilde{\mathfrak{u}}_1=\tilde{u}_x(x,t)\mid_{x=c(t)},
\label{DEb}
\\
\mathfrak{u}_2=u_t(x,t)\mid_{x=c(t)}, ~~\tilde{\mathfrak{u}}_2=\tilde{u}_t(x,t)\mid_{x=c(t)},
\label{DEc}
\\
\dot{\mathfrak{u}}=\frac{d u(c(t),t)}{dt}=c'(t)\mathfrak{u}_1+\mathfrak{u}_2,
~~
\dot{\tilde{\mathfrak{u}}}=\frac{d \tilde{u}(c(t),t)}{dt}=c'(t)\tilde{\mathfrak{u}}_1+\tilde{\mathfrak{u}}_2.
\label{DEd}
\end{eqnarray}
\label{DE}
\end{subequations}

\begin{claim}
The defect NLS equation (\ref{defectNLS}) can be described by the Lagrangian (\ref{LNLS}).
\end{claim}

Indeed, we consider the complete action
\begin{eqnarray}
\mathcal{A}=\int_{-\infty}^{\infty} dt\left\{\int_{-\infty}^{c(t)}dx \mathcal{L}(\tilde{u})+D+\int_{c(t)}^{\infty}dx \mathcal{L}(u)\right\}.
\label{aNLS}
\end{eqnarray}
The variation of $\mathcal{A}$ with respect to $u^*$ gives
\begin{eqnarray}
\delta\mathcal{A}=\int_{-\infty}^{\infty} dt\left\{\int_{c(t)}^{\infty}dx \left(\frac{\partial\mathcal{L}(u)}{\partial u^*}\delta u^*+\frac{\partial\mathcal{L}(u)}{\partial u^*_x}\delta u^*_x+\frac{\partial\mathcal{L}(u)}{\partial u^*_t}\delta u^*_t\right)+\frac{\partial D}{\partial \mathfrak{u}^*}\delta \mathfrak{u}^*+\frac{\partial D}{\partial \dot{\mathfrak{u}}^*}\delta \dot{\mathfrak{u}}^*\right\}.
\label{vauNLS}
\end{eqnarray}
Integrating the second term in (\ref{vauNLS}) by parts with respect to $x$, we find
\begin{eqnarray}
\begin{split}
\int_{-\infty}^{\infty} dt\int_{c(t)}^{\infty}dx \left(\frac{\partial\mathcal{L}(u)}{\partial u^*_x}\delta u^*_x\right)
&=-\int_{-\infty}^{\infty} dt\left\{\left.\left(\frac{\partial\mathcal{L}(u)}{\partial u^*_x}\delta u^*\right)\right|_{x=c(t)}
+\int_{c(t)}^{\infty}dx \left(\left(\frac{\partial\mathcal{L}(u)}{\partial u^*_x}\right)_x\delta u^*\right)\right\}
\\
&=\int_{-\infty}^{\infty} dt\left\{\mathfrak{u}_1\delta \mathfrak{u}^*
-\int_{c(t)}^{\infty}dx \left(\left(\frac{\partial\mathcal{L}(u)}{\partial u^*_x}\right)_x\delta u^*\right)\right\}.
\end{split}
\label{vauNLS2}
\end{eqnarray}
Using the identity
\begin{eqnarray}
\begin{split}
\frac{d}{dt}\left(\int_{c(t)}^{\infty}dx \left(\frac{\partial\mathcal{L}(u)}{\partial u^*_t}\delta u^*\right)\right)
&=-c'(t)\left.\frac{\partial\mathcal{L}(u)}{\partial u^*_t}\delta u^*\right|_{x=c(t)}
+\int_{c(t)}^{\infty}dx \left(\left(\frac{\partial\mathcal{L}(u)}{\partial u^*_t}\right)_t\delta u^*+\frac{\partial\mathcal{L}(u)}{\partial u^*_t}\delta u_t^*\right)
\\
&=\frac{i}{2}c'(t)\mathfrak{u}\delta \mathfrak{u}^*
+\int_{c(t)}^{\infty}dx \left(\left(\frac{\partial\mathcal{L}(u)}{\partial u^*_t}\right)_t\delta u^*+\frac{\partial\mathcal{L}(u)}{\partial u^*_t}\delta u_t^*\right),
\end{split}
\end{eqnarray}
the third term in (\ref{vauNLS}) can be written as
\begin{eqnarray}
\begin{split}
\int_{-\infty}^{\infty} dt\int_{c(t)}^{\infty}dx \left(\frac{\partial\mathcal{L}(u)}{\partial u^*_t}\delta u^*_t\right)
=-\int_{-\infty}^{\infty} dt\left\{\frac{i}{2}c'(t)\mathfrak{u}\delta \mathfrak{u}^*
+\int_{c(t)}^{\infty}dx \left(\left(\frac{\partial\mathcal{L}(u)}{\partial u^*_t}\right)_t\delta u^*\right)\right\}.
\end{split}
\label{vauNLS3}
\end{eqnarray}
Integrating the last term in (\ref{vauNLS}) by parts with respect to $t$, we have
\begin{eqnarray}
\int_{-\infty}^{\infty} dt\left(\frac{\partial D}{\partial \dot{\mathfrak{u}}^*}\delta \dot{\mathfrak{u}}^*\right)
=-\int_{-\infty}^{\infty} dt\left(\delta \mathfrak{u}^*\frac{d}{dt}\left(\frac{\partial D}{\partial \dot{\mathfrak{u}}^*}\right)\right).
\label{vauNLS5}
\end{eqnarray}
Inserting (\ref{vauNLS2}), (\ref{vauNLS3}) and (\ref{vauNLS5}) into (\ref{vauNLS}) and requiring the variation to be stationary, we obtain
\begin{eqnarray}
\begin{split}
0=&\int_{-\infty}^{\infty} dt\int_{c(t)}^{\infty}dx \left[\delta u^*\left(\frac{\partial\mathcal{L}(u)}{\partial u^*}-\frac{\partial}{\partial x}\left(\frac{\partial\mathcal{L}(u)}{\partial u^*_x}\right)-\frac{\partial}{\partial t}\left(\frac{\partial\mathcal{L}(u)}{\partial u^*_t}\right)\right)\right]
\\
&+\int_{-\infty}^{\infty} dt \left[\delta u^*\left(\mathfrak{u}_1-\frac{i}{2}c'(t)\mathfrak{u}+\frac{\partial D}{\partial \mathfrak{u}^*}-\frac{d}{dt}\left(\frac{\partial D}{\partial \dot{\mathfrak{u}}^*}\right)\right)\right].
\end{split}
\label{vauNLS0}
\end{eqnarray}
Similarly, requiring the variation of (\ref{aNLS}) with respect to $\tilde{u}^*$ to be stationary gives
\begin{eqnarray}
\begin{split}
0=&\int_{-\infty}^{\infty} dt\int_{-\infty}^{c(t)}dx \left[\delta \tilde{u}^*\left(\frac{\partial\mathcal{L}(\tilde{u})}{\partial \tilde{u}^*}-\frac{\partial}{\partial x}\left(\frac{\partial\mathcal{L}(\tilde{u})}{\partial \tilde{u}^*_x}\right)-\frac{\partial}{\partial t}\left(\frac{\partial\mathcal{L}(\tilde{u})}{\partial \tilde{u}^*_t}\right)\right)\right]
\\
&+\int_{-\infty}^{\infty} dt \left[\delta \tilde{u}^*\left(-\mathfrak{\tilde{u}}_1+\frac{i}{2}c'(t)\mathfrak{\tilde{u}}+\frac{\partial D}{\partial \mathfrak{\tilde{u}}^*}-\frac{d}{dt}\left(\frac{\partial D}{\partial \dot{\mathfrak{\tilde{u}}}^*}\right)\right)\right].
\end{split}
\label{vauNLS01}
\end{eqnarray}
Formulae (\ref{vauNLS0}) and (\ref{vauNLS01}) yield
\begin{subequations}
\begin{eqnarray}
\frac{\partial\mathcal{L}(u)}{\partial u^*}-\frac{\partial}{\partial x}\left(\frac{\partial\mathcal{L}(u)}{\partial u^*_x}\right)-\frac{\partial}{\partial t}\left(\frac{\partial\mathcal{L}(u)}{\partial u^*_t}\right)=0, \quad x>c(t),
\label{NLSa}
\\
\frac{\partial\mathcal{L}(\tilde{u})}{\partial \tilde{u}^*}-\frac{\partial}{\partial x}\left(\frac{\partial\mathcal{L}(\tilde{u})}{\partial \tilde{u}^*_x}\right)-\frac{\partial}{\partial t}\left(\frac{\partial\mathcal{L}(\tilde{u})}{\partial \tilde{u}^*_t}\right)=0, \quad x<c(t),
\label{NLSb}
\\
\mathfrak{u}_1-\frac{i}{2}c'(t)\mathfrak{u}+\frac{\partial D}{\partial \mathfrak{u}^*}-\frac{d}{dt}\left(\frac{\partial D}{\partial \dot{\mathfrak{u}}^*}\right)=0,
 \label{NLSda}
 \\
\tilde{\mathfrak{u}}_1-\frac{i}{2}c'(t)\tilde{\mathfrak{u}}-\frac{\partial D}{\partial \tilde{\mathfrak{u}}^*}+\frac{d}{dt}\left(\frac{\partial D}{\partial \dot{\tilde{\mathfrak{u}}}^*}\right)=0.
 \label{NLSdb}
\end{eqnarray}
\label{NLSd}
\end{subequations}
Equations (\ref{NLSa}) and (\ref{NLSb}) give nothing but (\ref{defectNLSa}) and (\ref{defectNLSb}),
while (\ref{NLSda}) and (\ref{NLSdb}), after some algebra, give exactly the defect conditions (\ref{defectNLSc}) and (\ref{defectNLSd}) at the defect location.

\subsubsection{Lagrangian formulation for the defect KdV equation}

For the KdV equation, the setting $u=q_{x}$ is suitable for a Lagrangian description.
In this setting, the defect KdV equation (\ref{defectkdv}) can be rewritten as
\begin{subequations}
\begin{eqnarray}
q_t+q_{xxx}+3q^2_{x}=0, ~~x>c(t),
\label{defectkdv1a}
\\
\tilde{q}_t+\tilde{q}_{xxx}+3\tilde{q}^2_{x}=0, ~~x<c(t),
\label{defectkdv1b}
\\
\left.\left(\tilde{q}_x+q_x\right)\right|_{x=c(t)}=-2\alpha-\frac{1}{2}\left.\left(\tilde{q}-q\right)^2\right|_{x=c(t)},
\label{defectkdv1c}
\\
\left.\left(\tilde{q}_t+q_t\right)\right|_{x=c(t)}=\left.\left[\left(\tilde{q}_{xx}-q_{xx}\right)\left(\tilde{q}-q\right)
-2\left((\tilde{q}_x)^2+(q_x)^2+\tilde{q}_xq_x\right)\right]\right|_{x=c(t)}.
\label{defectkdv1d}
\end{eqnarray}
\label{defectkdv1}
\end{subequations}
where $\alpha=-\frac{\beta^2}{4}$.
Regarding this defect system, we introduce the Lagrangian
\begin{eqnarray}
L=\int_{-\infty}^{c(t)}dx \mathcal{L}(\tilde{q})+D+\int_{c(t)}^{\infty}dx \mathcal{L}(q),
\label{Lkdv}
\end{eqnarray}
where
\begin{eqnarray}
\mathcal{L}(q)=\frac{1}{2}q_xq_t+(q_x)^3-\frac{1}{2}\left(q_{xx}\right)^2
\label{Lkdvr}
\end{eqnarray}
is the Lagrangian density of the bulk system for $x>c(t)$,
\begin{eqnarray}
\mathcal{L}(\tilde{q})=\frac{1}{2}\tilde{q}_x\tilde{q}_t+(\tilde{q}_x)^3-\frac{1}{2}\left(\tilde{q}_{xx}\right)^2
\label{Lkdvl}
\end{eqnarray}
is the Lagrangian density of the bulk system for $x<c(t)$,
\begin{eqnarray}
\begin{split}
D=&\frac{1}{4}\left(\mathfrak{q}\dot{\tilde{\mathfrak{q}}}-\tilde{\mathfrak{q}}\dot{\mathfrak{q}}\right)
-\frac{9}{40}\left(\tilde{\mathfrak{q}}-\mathfrak{q}\right)^5
-\left(\tilde{\mathfrak{q}}-\mathfrak{q}\right)^3\left(2\alpha+\frac{3}{4}\left(\tilde{\mathfrak{q}}_1+\mathfrak{q}_1\right)\right)
+\frac{1}{4}\left(\tilde{\mathfrak{q}}-\mathfrak{q}\right)^2\left(\tilde{\mathfrak{q}}_{11}-\mathfrak{q}_{11}\right)
\\
&-\left(\tilde{\mathfrak{q}}-\mathfrak{q}\right)\left((\tilde{\mathfrak{q}}_{1})^2+(\mathfrak{q}_{1})^2+\tilde{\mathfrak{q}}_{1}\mathfrak{q}_{1}
+3\alpha \left(\tilde{\mathfrak{q}}_{1}+\mathfrak{q}_{1}\right)+6\alpha^2\right)
+\frac{1}{2}\left(\tilde{\mathfrak{q}}_{11}-\mathfrak{q}_{11}\right)\left(\tilde{\mathfrak{q}}_{1}+\mathfrak{q}_{1}+2\alpha\right)
\\
&-c'(t)\left(\tilde{\mathfrak{q}}-\mathfrak{q}\right)\left(\alpha+\frac{1}{12}\left(\tilde{\mathfrak{q}}-\mathfrak{q}\right)^2 \right)
\end{split}
\label{Dkdv}
\end{eqnarray}
is the defect contribution at $x=c(t)$.
In (\ref{Dkdv}), we have used the following abbreviated expressions:
\begin{subequations}
\begin{eqnarray}
\mathfrak{q}=q(c(t),t), ~~\tilde{\mathfrak{q}}=\tilde{q}(c(t),t),
\label{DEakdv}
\\
\mathfrak{q}_1=q_x(x,t)\mid_{x=c(t)}, ~~\tilde{\mathfrak{q}}_1=\tilde{q}_x(x,t)\mid_{x=c(t)},
\label{DEbkdv}
\\
\mathfrak{q}_{11}=q_{xx}(x,t)\mid_{x=c(t)}, ~~\tilde{\mathfrak{q}}_{11}=\tilde{q}_{xx}(x,t)\mid_{x=c(t)},
\label{DEckdv}
\\
\mathfrak{q}_2=q_t(x,t)\mid_{x=c(t)}, ~~\tilde{\mathfrak{q}}_2=\tilde{q}_t(x,t)\mid_{x=c(t)},
\label{DEdkdv}
\\
\dot{\mathfrak{q}}=\frac{d q(c(t),t)}{dt}=c'(t)\mathfrak{q}_1+\mathfrak{q}_2,
~~
\dot{\tilde{\mathfrak{q}}}=\frac{d \tilde{q}(c(t),t)}{dt}=c'(t)\tilde{\mathfrak{q}}_1+\tilde{\mathfrak{q}}_2.
\label{DEfkdv}
\end{eqnarray}
\label{DEkdv}
\end{subequations}
In analogy to the case of NLS equation,
by requiring the variation of the complete action $\int_{-\infty}^{\infty} dt L$ to be stationary with respect to $q$ or $\tilde{q}$,
we find the following defect conditions:
\begin{subequations}
\begin{eqnarray}
\frac{1}{2}\dot{\mathfrak{q}}=-\frac{\partial D}{\partial \mathfrak{q}}+\frac{d}{dt}\left(\frac{\partial D}{\partial \dot{\mathfrak{q}}}\right),
~~\mathfrak{q}_{11}+\frac{\partial D}{\partial \mathfrak{q}_{1}}=0,
~~\frac{\partial D}{\partial \mathfrak{q}_{11}}=0,
 \label{kdvda}
 \\
\frac{1}{2}\dot{\tilde{\mathfrak{q}}}=\frac{\partial D}{\partial \tilde{\mathfrak{q}}}-\frac{d}{dt}\left(\frac{\partial D}{\partial \dot{\tilde{\mathfrak{q}}}}\right),
~~-\tilde{\mathfrak{q}}_{11}+\frac{\partial D}{\partial \tilde{\mathfrak{q}}_{1}}=0,
~~\frac{\partial D}{\partial \tilde{\mathfrak{q}}_{11}}=0.
 \label{kdvdb}
\end{eqnarray}
\label{kdvd}
\end{subequations}
Using (\ref{Dkdv}) the above defect conditions are exactly equivalent to (\ref{defectkdv1c}) and (\ref{defectkdv1d}).
To sum up, we find
\begin{claim}
The defect KdV equation (\ref{defectkdv1}) can be described by the Lagrangian (\ref{Lkdv}).
\end{claim}

\subsubsection{Lagrangian formulation for the defect mKdV equation}

For the mKdV equation in the setting $u=q_x$, an alternative Darboux matrix $B$  can be taken as
\begin{eqnarray}
B=I+\frac{i\beta}{2\lambda}\left( \begin{array}{cc} \cos(\tilde{q}-q) & -\sin(\tilde{q}-q) \\
 -\sin(\tilde{q}-q) &  -\cos(\tilde{q}-q) \\ \end{array} \right),
 \label{mkdv1dm}
\end{eqnarray}
and the corresponding BT becomes \cite{CZ2006}
\begin{eqnarray}
\begin{split}
\left(\tilde{q}_x+q_x\right)=\beta\sin(\tilde{q}-q),
\\
\left(\tilde{q}_t+q_t\right)=-\beta\left[ (\tilde{q}_{xx}-q_{xx})\cos(\tilde{q}-q)+(\tilde{q}^2_x+q^2_x)\sin(\tilde{q}-q)\right].
\end{split}
\label{demkdv1}
\end{eqnarray}
Then the time-dependent defect mKdV equation in the potential $q$ reads
\begin{subequations}
\begin{eqnarray}
q_t+q_{xxx}+2\left(q_{x}\right)^3=0, \quad x>c(t),
\label{defectmkdv1a}
\\
q_t+q_{xxx}+2\left(q_{x}\right)^3=0, \quad x<c(t),
\label{defectmkdv1b}
\\
\left.\left(\tilde{q}_x+q_x\right)\right|_{x=c(t)}=\beta\left.\sin(\tilde{q}-q)\right|_{x=c(t)},
\label{defectmkdv1c}
\\
\left.\left(\tilde{q}_t+q_t\right)\right|_{x=c(t)}=-\beta\left.\left[ (\tilde{q}_{xx}-q_{xx})\cos(\tilde{q}-q)+(\tilde{q}^2_x+q^2_x)\sin(\tilde{q}-q)\right]\right|_{x=c(t)}.
\label{defectmkdv1d}
\end{eqnarray}
\label{defectmkdv1}
\end{subequations}
We introduce the Lagrangian
\begin{eqnarray}
L=\int_{-\infty}^{c(t)}dx \mathcal{L}(\tilde{q})+D+\int_{c(t)}^{\infty}dx \mathcal{L}(q),
\label{Lmkdv}
\end{eqnarray}
where
\begin{eqnarray}
\mathcal{L}(q)=\frac{1}{2}q_xq_t+\frac{1}{2}(q_x)^4-\frac{1}{2}\left(q_{xx}\right)^2
\label{Lmkdvr}
\end{eqnarray}
is the Lagrangian density of the bulk system for $x>c(t)$,
\begin{eqnarray}
\mathcal{L}(\tilde{q})=\frac{1}{2}\tilde{q}_x\tilde{q}_t+\frac{1}{2}(\tilde{q}_x)^4-\frac{1}{2}\left(\tilde{q}_{xx}\right)^2
\label{Lmkdvl}
\end{eqnarray}
is the Lagrangian density of the bulk system for $x<c(t)$,
\begin{eqnarray}
\begin{split}
D=&\frac{1}{4}\left(\mathfrak{q}\dot{\tilde{\mathfrak{q}}}-\tilde{\mathfrak{q}}\dot{\mathfrak{q}}\right)
+\frac{1}{2}\left(\tilde{\mathfrak{q}}_{11}-\mathfrak{q}_{11}\right)
\left(\tilde{\mathfrak{q}}_{1}+\mathfrak{q}_{1}-\beta \sin(\tilde{\mathfrak{q}}-\mathfrak{q})\right)
\\
&+\frac{\beta}{6}\cos(\tilde{\mathfrak{q}}-\mathfrak{q})\left[\left(\tilde{\mathfrak{q}}_{1}\right)^2+\left(\mathfrak{q}_{1}\right)^2-4\tilde{\mathfrak{q}}_{1}\mathfrak{q}_{1}
+\beta(\tilde{\mathfrak{q}}_{1}+\mathfrak{q}_{1})\sin(\tilde{\mathfrak{q}}-\mathfrak{q}) +\beta^2\right]
-\frac{1}{2}\beta c'(t)\cos(\tilde{\mathfrak{q}}-\mathfrak{q})
\end{split}
\label{Dmkdv}
\end{eqnarray}
is the defect contribution at $x=c(t)$.
In (\ref{Dmkdv}) we have used the same abbreviated expressions as used in the case of defect KdV equation (see (\ref{DEkdv})).
In analogy to the case of defect KdV equation,
we find
\begin{claim}
The defect mKdV equation (\ref{defectmkdv1}) can be described by the Lagrangian (\ref{Lmkdv}).
\end{claim}

{\bf Remark 1.}
Taking $c'(t)=0$ in (\ref{LNLS}), (\ref{Lkdv}) and (\ref{Lmkdv}) respectively,
from our Lagrangian formulations for the time-dependent defect systems we can recover the corresponding Lagrangian formulations
for the defect systems in the situation of the defect being fixed \cite{CZ2006}.

\section{Integrability of the time-dependent defect system}

In this section, we will establish the integrability of the defect system both by constructing an infinite set of conserved densities
and by implementing the classical $r$-matrix method.
This analysis is based on an extension of the results of \cite{Caudrelier2008,CK2015}
from the situation of the defect being fixed to the situation of the defect moving with time.

\subsection{Conservation laws}
By generalizing the analogous result of \cite{Caudrelier2008}, we find the following conservation densities for the time-dependent defect system.
\begin{proposition}\label{pro1}
The generating function for the integrals of motion reads
\begin{eqnarray}
I(\lambda)=I_{bulk}^{left}(\lambda)+I_{bulk}^{right}(\lambda)+I_{defect}(\lambda),
\label{IM}
\end{eqnarray}
where
\begin{eqnarray}
I_{bulk}^{left}(\lambda)=\int_{-\infty}^{c(t)}\tilde{u}\tilde{\Gamma} dx,
\\
I_{bulk}^{right}(\lambda)=\int_{c(t)}^{\infty}u\Gamma dx,
\\
I_{defect}(\lambda)=-\left.\ln(B_{11}+B_{12}\Gamma)\right|_{x=c(t)},
\label{IMdefect}
\end{eqnarray}
and $B_{jk}$, $j,k=1,2$, is the $jk$-entry of the defect matrix $B$.
\end{proposition}
{\bf Proof}
From (\ref{CL}), we have
\begin{subequations}
\begin{eqnarray}
\left(u\Gamma\right)_t=\left(V_{11}+V_{12}\Gamma\right)_x, ~~x>c(t),
\\
\left(\tilde{u}\tilde{\Gamma}\right)_t=\left(\tilde{V}_{11}+\tilde{V}_{12}\tilde{\Gamma}\right)_x, ~~x<c(t),
\end{eqnarray}
\label{CL1}
\end{subequations}
where $\tilde{\Gamma}=\frac{\tilde{\phi}_2}{\tilde{\phi}_1}$.
Using (\ref{CL1}) and the rapid decay of the fields $u(x,t)$, $v(x,t)$, $\tilde{u}(x,t)$, $\tilde{v}(x,t)$, we find
\begin{eqnarray}
\left(\int_{-\infty}^{c(t)}\tilde{u}\tilde{\Gamma} dx+\int_{c(t)}^{\infty}u\Gamma dx\right)_t
=\left.\left(\tilde{V}_{11}+\tilde{V}_{12}\tilde{\Gamma}-V_{11}-V_{12}\Gamma\right)\right|_{x=c(t)}+c'(t)\left.\left(\tilde{u}\tilde{\Gamma}-u\Gamma\right)\right|_{x=c(t)}.
\label{der1}
\end{eqnarray}
From (\ref{BT}), we have
\begin{eqnarray}
\left.\tilde{\Gamma}\right|_{x=c(t)} =\left. \frac{B_{21}+B_{22}\Gamma}{B_{11}+B_{12}\Gamma}\right|_{x=c(t)}.
\label{gtildeg}
\end{eqnarray}
Using (\ref{rict}), (\ref{BT1b}) and (\ref{gtildeg}),  we obtain
\begin{eqnarray}
\left.\left(\tilde{V}_{11}+\tilde{V}_{12}\tilde{\Gamma}-V_{11}-V_{12}\Gamma\right)\right|_{x=c(t)}=\left.\frac{\left(B_{11}+B_{12}\Gamma\right)_t}{B_{11}+B_{12}\Gamma}\right|_{x=c(t)}.
\label{der11}
\end{eqnarray}
Using (\ref{ricx}), (\ref{BT1a}) and (\ref{gtildeg}),  we obtain
\begin{eqnarray}
\left.\left(\tilde{u}\tilde{\Gamma}-u\Gamma\right)\right|_{x=c(t)}=\left.\frac{\left(B_{11}+B_{12}\Gamma\right)_x}{B_{11}+B_{12}\Gamma}\right|_{x=c(t)}.
\label{der12}
\end{eqnarray}
Substituting (\ref{der11}) and (\ref{der12}) into (\ref{der1}), we obtain
\begin{eqnarray}
\left(\int_{-\infty}^{c(t)}\tilde{u}\tilde{\Gamma} dx+\int_{c(t)}^{\infty}u\Gamma dx\right)_t
=\left(\left.\ln(B_{11}+B_{12}\Gamma)\right|_{x=c(t)}\right)_t.
\label{der2}
\end{eqnarray}
Thus
\begin{eqnarray}
\left(I(\lambda)\right)_t=0.
\end{eqnarray}
This completes the proof. \QEDB

{\bf Remark 2.}
Proposition \ref{pro1} implies that the integrals of motion for the time-dependent defect system take a very similar form as those for the system with a
defect being fixed \cite{Caudrelier2008}.
However, the proof for proposition \ref{pro1} is technically more involved than analogous proof for the case that the defect being fixed
(see section 1.2 in \cite{Caudrelier2008});
we need to pay more attention to the $t$-derivatives of the associated quantities.

\subsection{Canonical transformation and classical $r$-matrix approach}

Canonical properties of BTs to integrable nonlinear evolution equations have been established in \cite{KW1976,K1977}.
Recently, by introducing a new Poisson bracket (called equal-space bracket),
it was shown in \cite{CK2015} that a defect condition described by a frozen BT can be interpreted naturally as a canonical transformation of the system.
As a consequence, the classical $r$-matrix approach \cite{Skly,Skly1988,Faddeev} can be implemented to establish Liouville integrability for the defect system
with a defect at a fixed location.
Here we show that analogous discussions can be adapted to the case of the time-dependent defect systems.

To fix ideas, we concentrate on the NLS equation.
Let us first recall some important results regarding the multi-symplectic formalism of the NLS equation \cite{CK2015}.
The key observation in \cite{CK2015} is to introduce
the following new equal-space Poisson bracket
\begin{eqnarray}
\begin{split}
\left\{u(x,t),u^*_x(x,\tau)\right\}=-\delta(t-\tau),
~~\left\{u_x(x,t),u^*(x,\tau)\right\}=\delta(t-\tau),
\\
\left\{u(x,t),u(x,\tau)\right\}=\left\{u(x,t),u^*(x,\tau)\right\}=\left\{u_x(x,t),u(x,\tau)\right\}=\left\{u_x(x,t),u_x(x,\tau)\right\}=0.
\end{split}
\label{NPB}
\end{eqnarray}
With this Poisson bracket, the NLS equation (\ref{NLS}) can be written in the following Hamiltonian form
\begin{eqnarray}
u_{xx}=\left\{u_x,H_T\right\},
\label{NLSHF}
\end{eqnarray}
where the new Hamiltonian $H_T$ are given by
\begin{eqnarray}
H_T=\int_{-\infty}^{\infty}d\tau \left(-|u_x|^2-\frac{i}{2}(u^*u_{\tau}-u_{\tau}^*u)+\varepsilon|u|^4\right).
\label{NLSH}
\end{eqnarray}
We construct a transition matrix from the time-part of the Lax pair:
Transition matrix $M_T(x,t, \lambda)$ is defined as the fundamental solution of the auxiliary linear problem (\ref{lpt}) with $M_T(x,-\infty, \lambda)=I$ (here $I$ denotes identity matrix),
\begin{eqnarray}
M_T(x,t, \lambda)=\overset{\curvearrowleft}\exp \int_{-\infty}^t V(x,\tau,\lambda)d\tau.
\end{eqnarray}
Using the Poisson bracket (\ref{NPB}), one can check directly that $M_T(x,t, \lambda)$ satisfies the following $r$-matrix relation \cite{CK2015}:
\begin{eqnarray}
\left\{M_{T1}(x,t,\lambda),M_{T2}(x,t,\mu)\right\}=\left[r(\lambda-\mu),M_T(x,t, \lambda)\otimes M_T(x,t, \mu)\right],
\label{rmatre}
\end{eqnarray}
where $M_{T1}(x,t,\lambda)=M_{T}(x,t,\lambda)\otimes I$, $M_{T2}(x,t,\mu)=I \otimes M_{T}(x,t,\mu)$, and
\begin{eqnarray}
r(\lambda-\mu)=\frac{\varepsilon}{2\left(\lambda-\mu\right)}\left( \begin{array}{cccc}
1 & 0 & 0 & 0
\\
0 &  0 & 1 & 0
\\
0 & 1 & 0 & 0
\\
0 &  0 & 0 & 1
 \\ \end{array} \right).
\end{eqnarray}
As an application, we can deduce that integrals of motion constructed from the trace of the monodromy matrix $M_T(x,\infty, \lambda)$ are in involution with respect to the Poisson bracket (\ref{NPB}). Thus, Liouville integrability of the NLS equation with respect to the Poisson bracket (\ref{NPB}) is proved.
For a system without a defect, the above argument for Liouville integrability is equivalent to the standard argument with respect to the usual (equal-time) Poisson bracket (see \cite{Faddeev} for details). The advantage of the above argument is that it can be applied to establish Liouville integrability of a system with a defect;
see \cite{CK2015} for the case of the NLS equation with a defect being fixed at $x=x_0$
and see the following discussions for the system with time-dependent defect.

We now adapt the arguments of \cite{KW1976,K1977} about canonical transformations to the above new Poisson bracket:
Transformation, which maps $u$ to $\tilde{u}$, is canonical if the following Pfaffian form is relative integrable invariant
\begin{eqnarray}
\int_{-\infty}^{\infty}dt\left(u^*_xdu+u_xdu^*\right)+H_Tdx.
\label{pff}
\end{eqnarray}
That is
\begin{eqnarray}
\int_{-\infty}^{\infty}dt\left(\tilde{u}^*_xd\tilde{u}+\tilde{u}_xd\tilde{u}^*\right)+\tilde{H}_Tdx=\int_{-\infty}^{\infty}dt\left(u^*_xdu+u_xdu^*\right)+H_Tdx-dW.
\label{pffi}
\end{eqnarray}
Here
\begin{eqnarray}
W(u,u^*,\tilde{u},\tilde{u}^*;x)=F(u,u^*,\tilde{u},\tilde{u}^*)-Ex
\end{eqnarray}
(with $E$ being a real constant) is called a generator of the transformation.
From (\ref{pffi}), we obtain the transformation equations:
\begin{eqnarray}
\begin{split}
u_x=\frac{\delta F}{\delta u^*},
~~u^*_x=\frac{\delta F}{\delta u},
\\
\tilde{u}_x=-\frac{\delta F}{\delta \tilde{u}^*},
~~\tilde{u}^*_x=-\frac{\delta F}{\delta \tilde{u}}.
\end{split}
\label{Transfe}
\end{eqnarray}
For the NLS equation, we find that $F$ can be taken as
\begin{eqnarray}
F=\int_{-\infty}^{\infty}dt \left(\frac{i}{2}\varepsilon \Omega \left(\frac{\tilde{u}_t-u_t}{\tilde{u}-u}-\frac{\tilde{u}^*_t-u^*_t}{\tilde{u}^*-u^*}\right)
+\frac{1}{3}\varepsilon \Omega^3 -\Omega\left(|u|^2+|\tilde{u}|^2+\varepsilon \alpha^2\right)+i\alpha\left(u\tilde{u}^*-u^*\tilde{u}\right)\right).
\end{eqnarray}
Then, the transformation equation (\ref{Transfe}) becomes nothing but the BT (\ref{deNLS}) of the NLS equation.
Hence, the defect condition for the NLS equation can be interpreted as a canonical transformation with respect to the Poisson bracket (\ref{NPB}).

We now turn to the implementation of the classical $r$-matrix approach to the NLS equation in the presence of the time-dependent defect.
For the time-dependent defect NLS system, we define the transition matrix as follows
\begin{eqnarray}
\begin{split}
\mathcal{M}(x,t,\lambda)=\left\{\begin{array}{l}
\widetilde{M}_T(x,t,\lambda), \quad -\infty<x<c(t),
\\
M_T(x,t,\lambda), \quad c(t)\leq x<\infty,
\end{array}\right.
\end{split}
\label{M1}
\end{eqnarray}
where $\widetilde{M}_T(x,t,\lambda)$ is the analogous matrix of $M_T(x,t,\lambda)$ but defined by the new canonical variable $\tilde{u}$.
Due to the canonical property of the transformation,
we immediately conclude that $\mathcal{M}(x,t,\lambda)$ satisfies the same $r$-matrix relation as that of $M_T(x,t,\lambda)$, that is
\begin{eqnarray}
\left\{\mathcal{M}_{1}(x,t,\lambda),\mathcal{M}_{2}(x,t,\mu)\right\}=\left[r(\lambda-\mu),\mathcal{M}(x,t, \lambda)\otimes \mathcal{M}(x,t, \mu)\right].
\label{rmatreM}
\end{eqnarray}
As a result, the trace of the monodromy matrix $\mathcal{M}(x,\infty,\lambda)$ provides a generating function for the conserved quantities
that are in involution with respect to the Poisson bracket (\ref{NPB}).
Thus, we establish Liouville integrability of the time-dependent defect NLS system (\ref{defectNLS}).

\section{The case of multiple time-dependent defects}

In this section, we generalize the above arguments for integrability to the case that there are multiple time-dependent defects in a system.

Let us first fix some notations.
We assume that $c_1(t)$, $c_2(t)$, $\cdots$, $c_n(t)$ are $n$ functions of class $C^1$ such that $c_1(t)<c_2(t)<\cdots<c_n(t)$.
We consider $n+1$ auxiliary problems for $\phi^{(j)}$, $j=0,\cdots,n$, with Lax pair
$U^{(j)}$, $V^{(j)}$ defined as in (\ref{LPxt}) with the fields $u^{(j)}$, $v^{(j)}$, replacing $u$, $v$.
We assume that the auxiliary problem for $U^{(0)}$, $V^{(0)}$ exists for $x<c_1(t)$,
the one for $U^{(j)}$, $V^{(j)}$ exists for $c_j(t)<x<c_{j+1}(t)$, $j=1,\cdots,n-1$,
and the one for $U^{(n)}$, $V^{(n)}$ exists for $x>c_n(t)$.
At $x=c_j(t)$, $j=1,\cdots,n$, we assume that the two systems are connected via the condition
\begin{eqnarray}
\phi^{(j-1)}(c_j(t),t,\lambda)=B^{(j)}(c_j(t),t,\lambda)\phi^{(j)}(c_j(t),t,\lambda),
\label{BTj}
\end{eqnarray}
where $B^{(j)}(x,t,\lambda)$, $j=1,\cdots,n$, satisfy
\begin{subequations}
\begin{eqnarray}
B^{(j)}_x(x,t,\lambda)=U^{(j-1)}(x,t,\lambda)B^{(j)}(x,t,\lambda)-B^{(j)}(x,t,\lambda)U^{(j)}(x,t,\lambda),
\label{BTj1a}
\\
B^{(j)}_t(x,t,\lambda)=V^{(j-1)}(x,t,\lambda)B^{(j)}(x,t,\lambda)-B^{(j)}(x,t,\lambda)V^{(j)}(x,t,\lambda).
\label{BTj1b}
\end{eqnarray}
\label{BTj1}
\end{subequations}
Given these notations, we find the following proposition, whose proof is similar to that of proposition \ref{pro1}.
\begin{proposition}
In the presence of multiple time-dependent defects, the generating function for the integrals of motion reads
\begin{eqnarray}
I(\lambda)=\int_{-\infty}^{c_1(t)}u^{(0)}\Gamma^{(0)} dx
+\sum_{j=1}^{n-1}\int_{c_{j}(t)}^{c_{j+1}(t)}u^{(j)}\Gamma^{(j)} dx
+\int_{c_n(t)}^{\infty}u^{(n)}\Gamma^{(n)} dx
+I_{defect}(\lambda),
\label{IMn}
\end{eqnarray}
where
\begin{eqnarray}
I_{defect}(\lambda)=-\sum_{j=1}^n\left.\ln(B^{(j)}_{11}+B^{(j)}_{12}\Gamma^{(j)})\right|_{x=c_j(t)}.
\label{IMdefectn}
\end{eqnarray}
Here $B^{(j)}_{11}$ and $B^{(j)}_{12}$  denote respectively the $11$-entry and $12$-entry of the defect matrix $B^{(j)}$.
\end{proposition}

The classical $r$-matrix approach can also be implemented to the system with multiple defects.
Indeed, we define the transition matrix for the defect NLS equation as follows
\begin{eqnarray}
\begin{split}
\mathcal{M}(x,t,\lambda)=\left\{\begin{array}{l}
M^{(0)}_T(x,t,\lambda), \quad -\infty<x<c_1(t),
\\
M^{(j)}_T(x,t,\lambda), \quad c_j(t)\leq x<c_{j+1}(t),\quad j=1,\cdots,n-1,
\\
M^{(n)}_T(x,t,\lambda), \quad c_n(t)\leq x<\infty,
\end{array}\right.
\end{split}
\label{M1}
\end{eqnarray}
where
\begin{eqnarray}
M^{(j)}_T(x,t, \lambda)=\overset{\curvearrowleft}\exp \int_{-\infty}^t V^{(j)}(x,\tau,\lambda)d\tau, \quad j=0,1,\cdots,n.
\end{eqnarray}
Then $\mathcal{M}(x,t,\lambda)$ satisfies the same $r$-matrix relation as that of (\ref{rmatreM}).
This fact immediately yields the Poisson commutativity of the motion integrals that generated from the trace of the monodromy matrix $\mathcal{M}(x,\infty,\lambda)$.

\section{Soliton solutions meeting the defects}

It is now our aim to seek soliton solutions that can meet the defect of a system.
To fix our ideas, we consider the KdV equation as an illustrative example.
We will focus on an interesting case: the defect moves at a constant speed.
We will show that the KdV equation with such a time-dependent defect admits peakon solutions.

Recall that a single-soliton  for the KdV equation is given by
\begin{eqnarray}
u(x,t)=\frac{8k^2}{\left(\alpha_1 \exp(\xi)+\alpha^{-1}_1\exp(-\xi)\right)^2}, \quad \xi=k(x-4k^2t), \quad k>0,
\label{s1kdv}
\end{eqnarray}
where $\alpha_1$ is a positive constant.
We assume that the defect takes place at $x=4k^2t$ (i.e. the speed of the defect is in coincidence with the wave speed).
In the presence of such a defect, we take the soliton on the other side of the defect in a similar form
\begin{eqnarray}
\tilde{u}(x,t)=\frac{8k^2}{\left(\alpha_2 \exp(\xi)+\alpha^{-1}_2\exp(-\xi)\right)^2}, \quad \xi=k(x-4k^2t), \quad k>0,
\label{s2kdv}
\end{eqnarray}
where $\alpha_2$ is a parameter to be determined by the defect condition.
By applying the defect condition (\ref{defectkdvc}) and (\ref{defectkdvd}), we find that $\alpha_2$ is determined by
\begin{eqnarray}
\begin{split}
2k \left(\frac{\alpha_1-\alpha_1^{-1}}{\left(\alpha_1+\alpha_1^{-1}\right)^3}+\frac{\alpha_2-\alpha_2^{-1}}{\left(\alpha_2+\alpha_2^{-1}\right)^3}\right)
=&\left(\frac{1}{\left(\alpha_1+\alpha_1^{-1}\right)^2}-\frac{1}{\left(\alpha_2+\alpha_2^{-1}\right)^2}\right)
\\
&\times \sqrt{\beta^2-16k^2\left(\frac{1}{\left(\alpha_1+\alpha_1^{-1}\right)^2}+\frac{1}{\left(\alpha_2+\alpha_2^{-1}\right)^2}\right)}.
\end{split}
\label{alpha2}
\end{eqnarray}
We further restrict our attention to find a solution such that there is no discontinuity at the defect.
This requirement implies that
\begin{eqnarray}
\left(\alpha_1+\alpha_1^{-1}\right)^2=\left(\alpha_2+\alpha_2^{-1}\right)^2.
\label{alpha12}
\end{eqnarray}
A nontrivial solution for $\alpha_2$ satisfying both (\ref{alpha2}) and (\ref{alpha12}) is that $\alpha_2=\alpha_1^{-1}$.
Let $\alpha_1=\exp\gamma$.
The solutions (\ref{s1kdv}) and (\ref{s2kdv}) on each side of the defect can be written in a uniform form:
$u(x,t)=2k^2 \sech^2\left(|\xi|+\gamma\right)$.
To sum up, we find
\begin{proposition}
Let the defect move at a constant speed $x=4k^2t$, and let the defect condition be defined by (\ref{defectkdvc}) and (\ref{defectkdvd}) with $x=c(t)$ replaced by $x=4k^2t$.
The KdV equation with such a defect admits the following single-peakon solution
\begin{eqnarray}
u(x,t)=2k^2 \sech^2\left(|\xi|+\gamma\right), \quad \xi=k(x-4k^2t).
\label{peakonkdv}
\end{eqnarray}
\end{proposition}

If $\gamma>0$, (\ref{peakonkdv}) presents a peakon wave with discontinuous first derivative at the peak (at $\xi=0$);
see figure 1 for a profile of this wave.
If $\gamma<0$, (\ref{peakonkdv}) presents a wave with two peaks (at $\xi=\pm\gamma$) pointing upwards and one peak (at $\xi=0$) pointing downwards
(called an anti-peakon simply),
where the first derivative is discontinuous at $\xi=0$ (the position of the anti-peakon); see figure 2 for a profile of this wave.
We note that the existence of peakon solutions was known as a typical feature of the CH type equations \cite{CH1993,FF1981}.
Here our results show that the usual soliton equations in the presence of time-dependent defects (such as the defect KdV equation discussed above)
can also admit peakon solutions.
We should emphasize that peakon solutions for the CH type equations and peakon solutions presented here should be interpreted in two different senses:
the former ones should be interpreted in a suitable weak sense,
while the latter ones should be interpreted in the sense that there is a time-dependent defect.

We now extend the above results to the case of the KdV equation with multiple defects located at different positions.
For clarity, we will assume, in the following, $\gamma>0$ and the defects located respectively at $\xi=j \gamma$, $j=0,\pm 1,\cdots,\pm N$.
As above, we restrict our attention to a solution which is continuous at the defects.
In analogy with the analysis used above, we find
\begin{proposition}
Assume that $\gamma>0$ and the defects locate at $\xi=j \gamma$, $j=0,\pm 1,\cdots,\pm N$.
Let the defect conditions be defined by (\ref{defectkdvc}) and (\ref{defectkdvd}) with $x=c(t)$ replaced by $x=4k^2t+jk^{-1} \gamma$, $j=0,\pm 1,\cdots,\pm N$.
The KdV equation with such multiple defects admits the following multi-peakon solution
\begin{eqnarray}
u(x,t)=\left\{\begin{array}{l} 2k^2 \sech^2\left(|\xi+N\gamma|+\gamma\right), \quad -\infty<\xi\leq -(N-1)\gamma,
\\
2k^2 \sech^2\left(|\xi+(N-2j)\gamma|+\gamma\right), \quad (2j-1-N)\gamma<\xi\leq (2j+1-N)\gamma, \quad j=1,\cdots,N,
\\
2k^2 \sech^2\left(|\xi-N\gamma|+\gamma\right), \quad (N+1)\gamma<\xi< \infty,
\end{array}\right.
\label{mpeakonkdv}
\end{eqnarray}
where $\xi=k(x-4k^2t)$.
\end{proposition}

The above solution (\ref{mpeakonkdv}) represents a wave which has $(N+1)$ peakons at $\xi=(2m-N)\gamma$, $m=0,1,\cdots N$,
and $N$ anti-peakons at $\xi=(2m-1-N)\gamma$, $m=1,\cdots N$.
For example, for $N=1$, it has two peakons at $\xi=\pm \gamma$, and one anti-peakon at $\xi=0$, and it looks like a ``M" shape wave;
see figure 3 for a profile of this M-shape wave solution.

\begin{figure}
\begin{minipage}[t]{0.3\linewidth}
\centering
\includegraphics[width=2.1in]{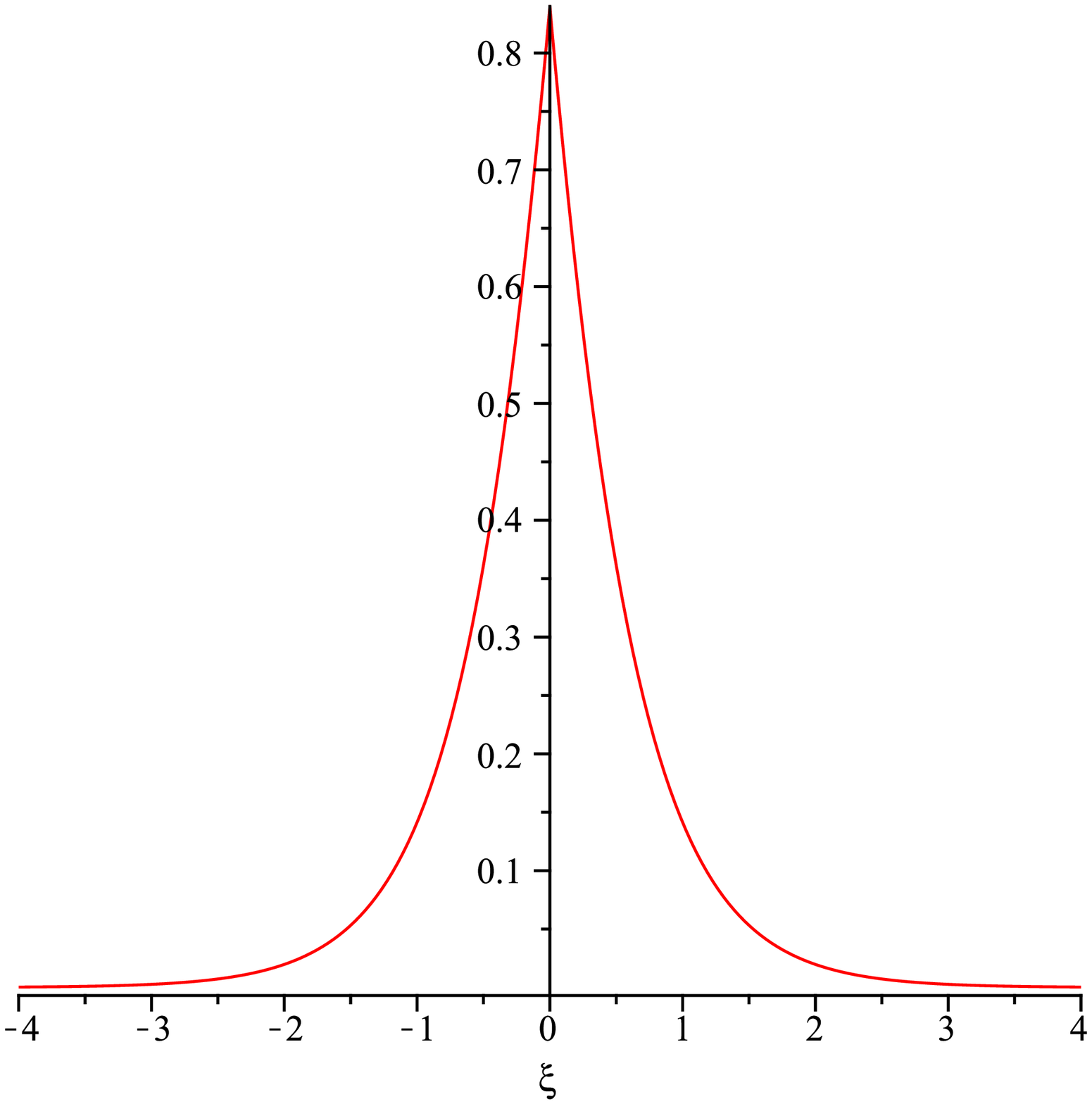}
\caption{\small{The single peakon solution determined by (\ref{peakonkdv}) with parameters $k=\gamma=1$.}}
\label{F21}
\end{minipage}
\hspace{1.9ex}
\begin{minipage}[t]{0.3\linewidth}
\centering
\includegraphics[width=2.1in]{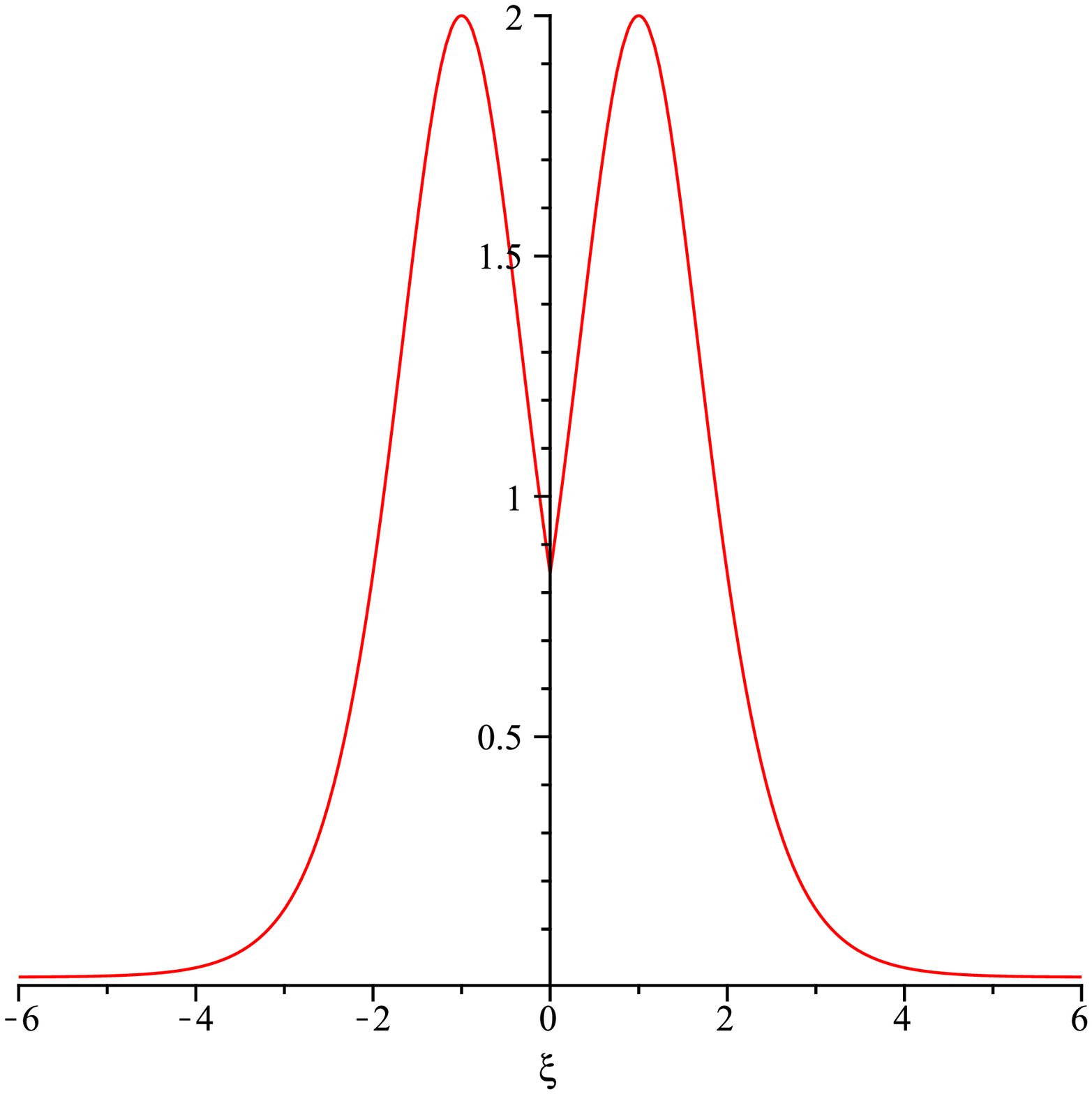}
\caption{\small{The solution determined by (\ref{peakonkdv}) with parameters $k=-\gamma=1$.}}
\label{F22}
\end{minipage}
\hspace{1.9ex}
\begin{minipage}[t]{0.3\linewidth}
\centering
\includegraphics[width=2.1in]{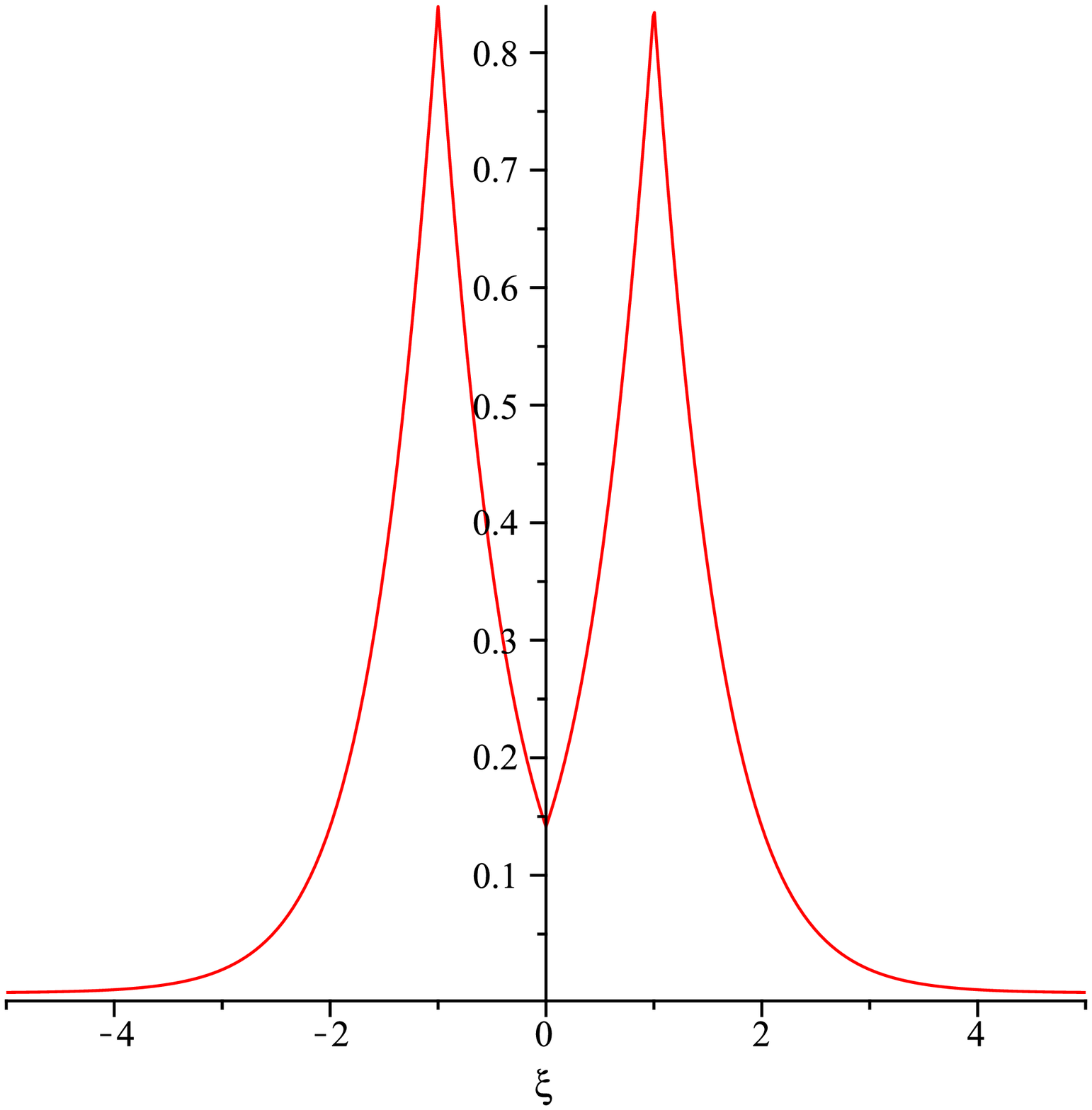}
\caption{\small{The M-shape peakon solution determined by (\ref{mpeakonkdv}) with $N=1$ and parameters $k=\gamma=1$.}}
\label{F23}
\end{minipage}
\end{figure}

{\bf Remark 3.} The above solutions are derived by a direct ansatz for the fields to either side of the defect tuned to satisfy the defect condition.
The fact that the defect condition is corresponding to a BT implies that we can systematically construct the solution of the defect system in the following way.
Given a solution $u(x,t)$ of the bulk system for $x\in \left(c(t),\infty\right)$,
we first implement a BT for all $x$, $t$ to find $\tilde{u}(x,t)$.
Then we define $\tilde{u}(x,t)$ as the solution of the bulk system for $x\in \left(-\infty,c(t)\right)$.
The solution constructed in such a manner solves the equation in the bulk as well as satisfies the defect condition at $x=c(t)$,
thus it provides a solution of the defect system.
For the case of the defect being fixed,
this strategy has been employed recently in \cite{CP2017} to construct finite-gap solutions for the defect KdV and sine-Gordon equations.
For the case of the defect moving with time as presented in this paper, similar considerations will be investigated in the future.

\section {Concluding remarks}

We have studied $(1+1)$-dimensional integrable soliton equations associated with the AKNS system in the presence of time-dependent defects.
We defined the defect condition as a B\"{a}cklund transformation evaluated at the time-dependent defect location.
We demonstrated that the resulting defect systems admit Lagrangian descriptions and established the integrability of the resulting defect systems
both by constructing an infinite set of conserved densities and by implementing the classical $r$-matrix method.
We also studied soliton solutions for the defect systems.
Although our results are presented for integrable soliton equations in continuous case,
it is clear that analogous results can be applied to integrable soliton equations in discrete case,
such as the integrable discrete NLS equation and the Toda lattice equation.

\section*{ACKNOWLEDGMENTS}
This work was supported by the National Natural Science Foundation of China (Grant Nos. 11771186 and 11671177).

\vspace{1cm}
\small{

}
\end{document}